\documentclass[a4paper,11pt]{article}
\usepackage{amsmath,amsfonts,amssymb,amsthm}
\usepackage[mathscr]{eucal}
\theoremstyle{plain}
\newtheorem{theorem}{Theorem}
\newtheorem{lemma}{Lemma}
\newtheorem{proposition}{Proposition}

\newtheoremstyle{note}{\topsep}{\topsep}{\slshape}{}{\scshape}{}{ }{}
\theoremstyle{note}

\newtheorem{remark}{Remark}
\theoremstyle{definition}
\newtheorem{definition}{Definition}
\numberwithin{equation}{section}
\numberwithin{theorem}{section}
\numberwithin{lemma}{section}
\numberwithin{proposition}{section}
\numberwithin{corollary}{section}
\numberwithin{remark}{section}

\newcommand\field{\mathbb}
\newcommand\R{\field{R}}
\newcommand\C{\field{C}}
\newcommand\vspan{\operatorname{span}}
\newcommand\rank{\operatorname{rank}}
\newcommand\rmd{\mathrm{d}\mspace{1mu}}
\newcommand\Dt{\frac{\mathrm{d}\phantom{t} }{\mathrm{d}\mspace{1mu} t}} 
\newcommand\pder[2]{\dfrac{\partial #1 }{\partial #2}} 
\newcommand\mtext[1]{\quad\text{#1}\quad}
\newcommand\defset[2]{\left\{{#1}\;\vert \;\; {#2} \,\right\}}
\title{Differential Galois obstructions for non-commutative integrability}
\author{Andrzej J.~Maciejewski,\\
  Institute of Astronomy,
  University of Zielona G\'ora \\
  Podg\'orna 50, PL-65--246 Zielona G\'ora, Poland,\\
  (e-mail: maciejka@astro.ia.uz.zgora.pl)       \and
  Maria Przybylska \\ 
Institut Fourier, UMR 5582 du CNRS,
Universit\'e de Grenoble I, \\\quad
100 rue des Maths,
BP 74, 38402 Saint-Martin d'H\`eres Cedex, France, and \\[0.5em]
Toru\'n Centre for Astronomy,
N.~Copernicus University,
Gagarina 11,\\\quad PL-87--100 Toru\'n, Poland,
(e-mail: mprzyb@astri.uni.torun.pl)
}

\begin{document}

\maketitle

\begin{abstract}
We show that if a holomorphic Hamiltonian system is holomorphically integrable
in the non-commutative sense in a neighbourhood of a non-equilibrium phase curve
which is located at a regular level of the first integrals, then the identity
component of the differential Galois group of the variational equations along
the phase curve is Abelian. Thus  necessary conditions for the commutative and
non-commutative integrability given by the differential Galois approach are the
same. 

\noindent MSC 37J30, 70H06, 53D20 
\end{abstract}

\section{Introduction}
One of the main problems of Hamiltonian mechanics is to decide whether a given
system is
integrable or not.   There exist only few methods which give effective and
rigorous necessary conditions for integrability. One of them arose from a
great idea of S.~N.~Kovalevskaya. It relates integrability with the properties
of solutions as a function of the complex time. Too complicated branching of
solutions is not compatible with the integrability. The idea of Kovalevskaya
was investigated for almost a century by leading mathematicians and physicists
of the epoch. Finally, at the beginning of eighties of the previous century, the
problem of mysterious relations between the integrability and branching of
solutions was explained by S.~L.~Ziglin in his elegant and powerfull theory
formulated in \cite{Ziglin:82::a,Ziglin:83::a}. In the Ziglin theory the
integrability is connected with the properties of the monodromy group of the
variational equations along a particular non-equilibrium solution of the
considered system. If the system is integrable, then the monodromy group cannot
be too `big'.  The Ziglin theory was successfully applied  to study the 
integrability of many
Hamiltonian systems, see e.g. \cite{Churchill:96::b,Yoshida:87::a,Yoshida:88::a}
and references therein.
In the middle of nineties of the previous century, thanks to the works of
A.~Baider,
R.~.C.~Churchill, J.~J.~Morales, J.-P.~Ramis, D.~L.~Rod, C.~Simo and
M.~F.~Singer  the Ziglin theory was considerably developed, see
\cite{Churchill:96::b,Morales:99::c,Morales:99::b,Morales:01::b1} and references
therein. The main idea of this extension  is  to use the
differential Galois group of the variational equations in order to obtain the
necessary conditions for the integrability.  We describe this approach shortly.
For a more detailed description see the cited papers.

Let $(M,\omega)$ be a $2n$-dimensional complex connected analytic symplectic
manifold. The symplectic form $\omega$ induces the Poisson bracket
$\{\cdot,\cdot\}$ and the corresponding Poisson tensor $\Lambda$ on $M$.  For a 
meromorphic function $H:M\rightarrow\C$,  
 we denote by $X_H$ the Hamiltonian vector field generated by $H$.  We have the
well known identities
\begin{equation}
 \{F,H\}=\omega(X_F,X_H)=\Lambda(\rmd F, \rmd H), 
\end{equation} 
for arbitrary meromorphic functions $F$ and $H$ on $M$. 
Consider the Hamiltonian equations
\begin{equation}
 \label{eq:hds}
\Dt x = X_H(x),  \qquad x\in M, \quad t\in\C. 
\end{equation} 
A meromorphic function $F:M\rightarrow\C$ is a first integral of
system~\eqref{eq:hds} iff 
$\{H,F\}=0$.  The Poisson bracket $\{F_1,F_2\}$ of two first integrals $F_1$ and
$F_2$ of~\eqref{eq:hds} is a first integral. Hence the set of all meromorphic
first integrals of~\eqref{eq:hds} is a Lie algebra with respect to the Poisson
bracket. 

A Hamiltonian system~\eqref{eq:hds} is meromorphically integrable in the
Liouville sense iff it
admits $n$ meromorphic first integrals which are functionally independent on an
open and dense subset of $M$. 

Let $\varphi(t)$ be a non-equilibrium particular solution of~\eqref{eq:hds}.
Here we consider $\varphi$ as a full analytic function, i.e., $\varphi$ is a
maximal analytic continuation of a local solution. It
defines a Riemann surface $\Gamma$ with $t$ as a local coordinate. 
The variational equations along $\Gamma$ have the form 
\begin{equation}
 \label{eq:svar}
\Dt y= A(t)y,  \qquad A(t) = \pder{X_H}{x}(\varphi(t)),\qquad y\in T_\Gamma M.
\end{equation} 
The coefficients of this equation belong to the field $\mathscr{M}(\Gamma)$ of
functions meromorphic on $\Gamma$. This is a differential field with the
derivative with respect to the time as the derivation. Its subfield of constants
is
$\C$. Let $\mathscr{G}$ denote the differential Galois group of
system~\eqref{eq:svar}.  It is an algebraic subgroup of $\mathrm{Sp}(2n,\C)$. 

The following theorem states that the existence of a commutative $n$-dimensional
Lie algebra of
functionally independent first integrals implies the commutativity of the
identity component of the differential Galois group of the variational
equations. It was formulated by J.~J.~Morales and J.~P.~Ramis
in~\cite{Morales:99::c,Morales:01::b1}, see also~\cite{Audin:01::c,Audin:02::c}. 
\begin{theorem}
 \label{thm:mora}
If Hamiltonian system~\eqref{eq:hds} possesses $n$ commuting functionally
independent meromorphic first integrals in a connected neighbourhood of a
non-equilibrium phase curve $\Gamma$, then the identity component of the
differential Galois group of the variational equations along $\Gamma$ is
Abelian.
\end{theorem}
Let us note that the first integrals are functionally independent in the neighbourhood of $\Gamma$ but not necessarily independent on
$\Gamma$ itself. 

The above theorem was successfully applied for proving the non-integrability of
many
systems, 
see e.g. 
\cite{Morales:99::b,Morales:01::b2,Audin:03::b,Audin:03::c,Tsygvintsev:01::a,Maciejewski:05::b,
Maciejewski:05::a}.
Its strength lies in two facts. The differential Galois group is bigger than the
monodromy group. Moreover, it is easier to determine the   differential Galois
group of given equations than their monodromy group.

In many cases Hamiltonian system~\eqref{eq:hds} on a $2n$ dimensional manifold
admits more than $n$ functionally independent but non-commuting first integrals,
see e.g. \cite{Brailov:83::,Brailov:86::,Adams:88::}.
Under certain conditions integration of such systems can be reduced to
quadratures. Such examples gave motivation for introducing the notion of
non-commutative integrability, see e.g.
\cite{Mishchenko:78::,Guillemin:83::,MR1972543,MR973403} and
references therein.  Let us remind shortly the idea  of  the  non-commutative
integrability. As we work with complex Hamiltonian systems, we adopt basic
definitions from \cite{MR1972543} to this context.

Let $F_i:M\rightarrow \C$  for $1\leq i\leq k$, be functions  holomorphic  in a
neighbourhood of a point  $x \in M$.  They define a natural map
\begin{equation}
\label{eq:mm}
 F:U\rightarrow\C^k, \qquad U\ni x\rightarrow F(x)=(F_1(x),\ldots,F_k(x))\in
\C^k.
\end{equation} 
In the cotangent space $T^\star_xM$ we distinguish a linear subspace $F_x$
spanned by differentials 
$\rmd F_i(x): T_xM\rightarrow\C$, i.e.
\begin{equation}
\label{eq:Fx}
 F_x := \vspan_\C\{\rmd F_1(x), \ldots,\rmd F_k(x) \}.
\end{equation} 
The Poisson tensor $\Lambda_x:=\Lambda(x)$ at point $x$ is  a bilinear form on
$T^\star_xM$ which induces a linear map $\Lambda_x^\sharp: T^\star_xM\rightarrow
T_xM$, defined by 
 \begin{equation}
\label{eq:Lsharp}
  \Lambda_x(u,v)=\langle u,\Lambda_x^\sharp (v)\rangle= u\cdot\Lambda_x^\sharp
(v)  := u\left( \Lambda_x^\sharp (v)\right), \mtext{for all} u,v \in T^\star_xM.
 \end{equation} 
To simplify the notation we write  $\Lambda_{|F_x}$ and
${\Lambda^\sharp}_{|F_x}$ to denote the restriction of $\Lambda_x$ and 
$\Lambda_x^\sharp$ to $F_x$, respectively.
\begin{definition}
\label{def:non}
We say that a holomorphic Hamiltonian system~\eqref{eq:hds} is holomorphically
integrable in the
non-commutative sense iff there exist $k$ holomorphic first integrals $F_1,
\ldots, F_k$ which are
functionally independent on an open and dense subset $U$ of $M$, and satisfy
\begin{equation}
\label{eq:FiFj}
 \{F_i,F_j\}(x)=a_{ij}(F_1(x),\ldots,F_k(x)), \mtext{for}x\in U, \mtext{and}
1\leq i,j\leq k. 
\end{equation} 
where $a_{ij}:\C^k\supset F^{-1}(U)\rightarrow \C $ are holomorphic functions;
and, moreover,  condition
\begin{equation}
\label{eq:dims}
 \dim_{\C} F_x  + \dim_{\C} \ker {\Lambda^\sharp}_{|F_x} = 2n \mtext{for}x\in U.
\end{equation}
is fullfilled. 
\end{definition}
Let us make some remarks about the above definition. 
\begin{remark}
 As it is assumed that functions $F_1, \ldots, F_k$ are functionally independent
at $x\in U$ we have $\dim_{\C} F_x=k$. Hence differentials $\rmd F_1(x), \ldots,
\rmd F_k(x) $ form a linear base in  $T^\star_xM$.  Let $v_x^1, \ldots, v_x^k\in
T_xM$  be the dual base. The matrix $A=[A_{ij}]$ of $\Lambda_{|F_x}$ is given
by 
\begin{equation}
\label{eq:Aij}
 A_{ij} := \Lambda_{|F_x}(\rmd F_i(x), \rmd F_j(x))=\{F_i,F_j\}(x)
=a_{ij}(F(x)).
\end{equation} 
 Let $B=[B_{ij}]$ be the matrix of ${\Lambda^\sharp}_{|F_x} $ in the chosen
bases, i.e.,
\begin{equation}
 {\Lambda^\sharp}_{|F_x} (\rmd F_j(x)) := \sum_{l=1}^k B_{lj}v_x^l. 
\end{equation}
Then, from~\eqref{eq:Aij} and~\eqref{eq:Lsharp} we obtain 
\begin{equation}
 A_{ij} = \Lambda_{|F_x}(\rmd F_i(x), \rmd F_j(x))= \langle \rmd F_i(x),
{\Lambda^\sharp}_{|F_x}(\rmd F_j(x)) \rangle = \sum_{l=1}^k B_{lj}  \langle \rmd
F_i(x),v_x^l \rangle= B_{ij}.
\end{equation} 
This shows that $A=B=[a_{ij}(F(x))]$. Hence 
\begin{equation}
 \dim_{\C} \ker {\Lambda^\sharp}_{|F_x}= k - \rank [a_{ij}(F(x))].
\end{equation}   
\end{remark}
\begin{remark}
 One can consider also  the non-commutative integrability with meromorphic first
integrals. We restrict ourselves to holomorphic first integrals to avoid
technical difficulties in the proof of our main results. 
\end{remark}
\begin{remark}
The definition of the non-commutative integrability for real Hamiltonian systems
can be obtained from the definition given above if we change $\C$ to $\R$. It
should be mentioned, however, that usually for real Hamiltonian systems it is
assumed that the first integrals are of class $C^\infty$.  
\end{remark}
\begin{remark}
 If  a real Hamiltonian system is integrable in the non-commutative sense
with   smooth  first integrals, then  the compact connected common levels of
these first integrals are
tori of dimension  smaller than $n$. 
\end{remark}

The well known conjecture of A.~T.~Fomenko and A.~S.~Mishchenko says that if a
system is
integrable in the non-commutative sense with the first integrals of a given
class,
then it is integrable in the usual Liouville sense with the first integrals
belonging to the same class of functions. Recently S.~T.~Sadetov
\cite{MR2120177} proved this
conjecture  for a case when the first integrals span a finite dimensional Lie
algebra, i.e. when 
\begin{equation*}
 \{F_i,F_j\}=\sum_{m=1}^k c_{ij}^m F_m,  
\end{equation*}
where $c_{ij}^m$ are constant. Furthermore in \cite{MR1972543} it was
shown that if a real Hamiltonian system is integrable in the non-commutative
sense,  then it is
integrable in the Liouville sense with the first integrals 
$F_1,\ldots,F_n\in\mathrm{C}^\infty(M)$.  
This important result shows, on the one hand, that the non-commutative
integrability `means'  almost the commutative one. But, on  the other hand,
still there remains an open problem. In fact, let us assume that the considered system
is
integrable in the non-commutative sense with the first integrals which are real
holomorphic. Does it imply that it is integrable in the Liouville sense with
real holomorphic first integrals? There is a conjecture in \cite{MR1972543} that
the answer to this question is affirmative.

The purpose of this paper is to give  necessary conditions for the
non-commutative
integrability of complex Hamiltonian systems.  More precisely, our aim is to
give  necessary conditions in the spirit of the Morales-Ramis
Theorem~\ref{thm:mora}.  The above mentioned results relating the
non-commutative
and commutative integrability suggest  that these necessary conditions should be
the same as for the commutative integrability.  Our main result shows that in
fact they are really  the same.  
\begin{theorem}
 \label{thm:my}
If Hamiltonian system~\eqref{eq:hds} is  integrable in the non-commutative sense
with   first integrals   holomorphic in a connected neighbourhood of a
non-equilibrium
phase curve $\Gamma$,  then the
identity component of the differential Galois group of the variational equations
along $\Gamma$ is Abelian.
\end{theorem}
The rest of this paper contains a proof of the above theorem. 

\section{Proof of Theorem~\ref{thm:my}}

Our proof is based on two facts. The first of them is the Lie-Cartan theorem
(see page 126 in \cite{Cartan:71::} and Section~3.2.2 of \cite{Arnold:97::})  which we formulate in a form adequate
for complex Hamiltonian systems. 

Let $(M,\omega)$ be a $2n$-dimensional complex connected analytic symplectic
manifold.   Assume that $F_1, \ldots, F_k$ are functionally independent
holomorphic 
functions defined in a non-empty connected open subset $W\subset M$, and 
\[
 F:  W\rightarrow \C^k, \qquad W \ni x\mapsto(F_1(x),\ldots,F_k(x))\in\C^k,
\]
is the momentum map. Moreover, we assume that 
 there exist holomorphic functions $a_{ij}:\C^k\rightarrow\C$ such that 
$\{F_i,F_j\}= a_{ij}\circ F$, i.e.
\[
 \{F_i,F_j\}(x)= a_{ij}(F_1(x),\ldots, F_k(x)), \mtext{for} i,j=1, \ldots, k.
\]

\begin{theorem}[Lie-Cartan]
 \label{thm:LC}
 Let $c=F(p)$, $p\in W$  and assume that the rank of matrix
$[a_{ij}]$ is constant in a neighbourhood of $c$. Then there exists a
neighbourhood $U\subset \C^k$ of point $c$ and $k$ functionally independent
holomorphic functions  $g_i:U\rightarrow \C$, such that functions 
\begin{equation}
 \label{eq:Gi}
G_i = g_i\circ F : M\supset F^{-1}(U)\rightarrow \C , \qquad i=1,\ldots, k,
\end{equation}
satisfy 
\[
 \{G_{2i-1},G_{2i}\}=1 \mtext{for} i = 1,\ldots, r,
\]
where $2r$ is the rank of $k\times k$ matrix $[a_{ij}]$. Moreover, all remaining
Poisson brackets of functions $G_i$ vanish. 
\end{theorem}
\begin{remark}
 To have an idea  how to prove the above theorem, let us restrict $W$ in
such a way that $\rank [a_{ij}(y)] = 2r$ for all $y\in P=F(W)$.  Then $P$ is an
analytic submanifold of $\C^k$. 
We can define a Poisson bracket $\{\cdot, \cdot\}_A$ on $P$ demanding 
\begin{equation}
 \{y_i,y_j\}_A:=a_{ij}(y_1,\ldots,y_k), \mtext{for} 1\leq i,j\leq k.
\end{equation} 
Thus, $P$ is a Poisson manifold. For a point $c\in P$ we can apply the local
structure theorem for a Poisson bracket, see e.g., p. 348 in
\cite{Marsden:99::}. From this theorem it follows that there exists a
neighbourhood $U$ of $c$, and holomorphic functions $g_i:U\rightarrow\C$, $1\leq
i \leq k$, such that 
\[
 \{g_{2i-1},g_{2i}\}_A=1 \mtext{for} i = 1,\ldots, r,
\]
and all the remaining brackets vanish.  Now, for functions~\eqref{eq:Gi} we have
\begin{equation}
 \{G_i,G_j\}(x)=a_{ij}(g_1(F(x)),\ldots,g_k(F(x)))=\{g_i,g_j\}_A(F(x))
\mtext{for}x\in F^{-1}(U).
\end{equation} 
\end{remark}
\begin{remark}
\label{rem:non}
Let point $c=(c_1,\ldots, c_k)\in \C^k$ and functions $G_i$, $1\leq i\leq k$ be
as in the above theorem. Moreover, we assume that $ k = n+r$. The common level 
\begin{equation*}
 \Sigma_c:=\defset{x\in W}{ G_i(x)=c_i, \quad 1\leq i\leq k },
\end{equation*}
is an analytic submanifold of $M$ and $\dim \Sigma_c = 2n -k= n -r$.   The
tangent space $T_x\Sigma_c$ to this manifold at
point $x$ is the intersection of kernels  of  differentials of all functions
$G_i$ at point $x$,  i.e.
\begin{equation}
 T_x\Sigma_c = \bigcap_{i=1}^k\defset{v\in T_x M}{\rmd G_i(x)\cdot v =0}=
\bigcap_{i=1}^k \ker \rmd G_i(x).
\end{equation} 
Note that   $k-2r=n-r$ functions  $G_{2r+1}, \ldots, G_k$ commute with  all 
other
 $G_1, \ldots, G_k$.    Now, denote by $X_i$ the Hamiltonian vector fields
generated by $G_i$ for $1\leq i\leq k$.  As functions $G_i$ are functionally
independent, these vector fields are linearly independent. We have
\begin{equation}
 0 = \{G_i,G_j\}(x)= \rmd G_i(x)\cdot X_j(x) \mtext{for}1\leq i\leq k,
\mtext{and}2r<j\leq k.
\end{equation} 
Hence $X_j(x)\in T_x\Sigma_c$ for $2r<j\leq k$, so these vector fields form a
linear base in 
$T_x\Sigma_c$. As
\begin{equation}
 0 = \{G_i,G_j\}(x)=\omega_x(X_i,X_j) \mtext{for}2r<i,j\leq k, 
\end{equation} 
$T_x\Sigma_c$ is an isotropic subspace of $T_xM$.
\end{remark}
To formulate and prove the second fact we consider   the symplectic Lie algebra
$\mathrm{sp}(V,\Omega)$ where $V$ is a complex symplectic vector space of
dimension $2n$ with a symplectic form $\Omega$. By definition,  elements of
$\mathrm{sp}(V,\Omega)$ are endomorphisms  $A:V\rightarrow V$ such that
$\Omega(x,Ay)=-\Omega(Ax,y)$ for all $x,y\in V$.   For an element $A \in
\mathrm{sp}(V,\Omega)$ we define a function 
\begin{equation*}
 H_A:V\rightarrow \C, \quad V\ni x \mapsto H_A(x)=\frac{1}{2}\Omega(Ax,x), 
\end{equation*}
and a vector field 
\begin{equation*}
 V\ni x \mapsto (x,v_A(x))\in TV \mtext{where} v_A(x)=Ax.
\end{equation*}
For a holomorphic function $F:V\rightarrow \C$ we denote  by $X_F$ the
corresponding Hamiltonian vector field.    
\begin{proposition}
 Let $A\in \mathrm{sp}(V,\Omega)$. Then $v_A = X_{H_A}$.
\end{proposition}
For a proof see Proposition~2.5.1 on page 77 in \cite{Marsden:99::}. 
\begin{proposition}
 Let $A,B\in \mathrm{sp}(V,\Omega)$. Then 
\begin{equation}
\label{prop:com}
 A\circ B = B\circ A \quad \Leftrightarrow \quad \{H_A,H_B\}=0 \quad
\Leftrightarrow \quad [v_A,v_B]=0.
\end{equation} 
\end{proposition}
An easy proof of the above statement we leave to the reader. As a matter of fact,
one can show that Lie algebra $ \mathrm{sp}(V,\Omega)$ is isomorphic to the Lie
algebra of linear Hamiltonian vector fields with  the commutator of vector
fields as the Lie bracket, and it is also isomorphic to the Lie algebra of
quadratic homogeneous Hamiltonian functions with the Poisson bracket as the Lie
bracket, see Section~3.4 in \cite{Morales:99::c}. 

Let   $U$ be a non-empty open subset $V$. Functions holomorphic on $U$ form a
ring denoted by $\mathscr{O}(U)$.
\begin{definition}
 We say that Lie algebra $\mathfrak{g}\subset \mathrm{sp}(V,\Omega)$ preserves a
function
$F\in \mathscr{O}(U) $ iff  $v_A[F]= \{F,H_A\}=0$ for all $A\in\mathfrak{g}$. 
\end{definition}
In other words, $\mathfrak{g}$ preserves $F$ iff $F$ is a common first integral
of all elements of $\mathfrak{g}$  considered as linear Hamiltonian vector
fields. 

The lemma below is a generalisation of the so called  Key Lemma, see \cite[Lemma
III.3.7, p.72]{Audin:01::c}.  
\begin{lemma}
 \label{lem:cle}
Assume that $F_1, \ldots, F_{n+r}\in\mathscr{O}(U)$, $0\leq r\leq n$ are
functionally independent on $U$ and 
\begin{equation}
\label{eq:fij}
 \{F_i,F_j\}= 0 \mtext{for} 1\leq i\leq n-r \mtext{and} 1\leq j\leq n+r.
\end{equation}
If a Lie algebra $\mathfrak{g}\subset \mathrm{sp}(V,\Omega)$ preserves all 
$F_j$ for $1\leq j \leq n+r$, then $\mathfrak{g}$ is Abelian. 
\end{lemma}
\begin{proof}
Let $X_j$ be the Hamiltonian vector field generated by $F_j$ for $1\leq j \leq
n+r$.  From~\eqref{eq:fij} we have 
\begin{equation}
\label{eq:xjxi}
 [X_j,X_i]=-X_{\{F_j,F_i\}} = 0  \mtext{for} 1\leq i\leq n-r \mtext{and} 1\leq
j\leq n+r.
\end{equation} 
Let $A\in \mathfrak{g}$ and $v_A$ be the corresponding linear Hamiltonian vector
field with Hamiltonian function $H_A$. Then 
\begin{equation}
\label{eq:xjva}
 [X_j,v_A]=-X_{\{F_j,H_A\}}=0, \mtext{for} 1\leq j\leq n+r, 
\end{equation} 
because $\{F_j,H_A\}=0 $  by assumption that $ \mathfrak{g}$ preserves  $F_j$
for $1\leq j\leq n+r$. 
 
 As functions $F_1, \ldots, F_{n+r}\in\mathscr{O}(U)$ are functionally
independent on $U$, their common level 
\begin{equation}
 \Sigma_c:=\defset{x\in U}{F_i(x)=c_i, \quad i =1,\ldots, n+r}, \qquad
c=(c_1,\ldots, c_k)\in\C^k,
\end{equation} 
is a $(n-r)$-dimensional  submanifold of $V$.  The tangent space $T_x  \Sigma_c$
to  $ \Sigma_c$ at point $x$ is the intersection 
\begin{equation}
  T_x\Sigma_c = \bigcap_{i=1}^{n+r}\defset{v\in T_x V}{\rmd F_i(x)\cdot v =0}=
\bigcap_{i=1}^{n+r} \ker \rmd F_i(x),
\end{equation} 
see Remark~\ref{rem:non}.   We show that  $X_1(x), \ldots, X_{n-r}(x) \in T_x 
\Sigma_c$.  In fact,  by~\eqref{eq:fij}, for arbitrary $1 \leq j\leq n+r$ and $1
\leq i\leq n-r$, equality
\begin{equation}
\rmd F_j(x)\cdot X_i(x)  =\{F_j,F_i\}(x) = 0, 
\end{equation} 
holds. Vector fields $X_1, \ldots, X_{n+r}$ are linearly independent at all
points of $\Sigma_c$. 
Hence we have $n-r$ linearly independent vector fields $X_1(x), \ldots, X_{n-r}(x)
\in T_x  \Sigma_c$. They form a linear base of $T_x  \Sigma_c$.

Now, let $A \in \mathfrak{g}$.  By assumption  $\mathfrak{g}$ preserves all
functions $F_j$, $1 \leq j\leq n+r$, so vector field $v_A(x)$ is tangent to
$\Sigma_c$ at $x$.  In fact, we have
\begin{equation}
 0=v_A[F_j](x)= \rmd F_j(x)\cdot v_A(x)=\{F_j,H_A\}(x), \mtext{for} 1\leq j\leq
n+r.
\end{equation} 
Hence
\begin{equation}
\label{eq:lin}
 v_A(x)= \sum_{i=1}^{n-r}\lambda_i(x) X_i(x),
\end{equation}  
because vector fields $X_1(x), \ldots, X_{n-r}(x)$ span $T_x\Sigma_c$.  In the
above formulae $\lambda_i$ are holomorphic functions.  Now, using~\eqref{eq:lin}
and~\eqref{eq:xjva},  we obtain
\begin{equation}
 0=[ X_j,v_A](x)=[ X_j, \sum_{i=1}^{n-r}\lambda_i X_i](x)=\sum_{i=1}^{n-r}\left(
X_j[\lambda_i] (x) X_i (x)+\lambda_i(x)[X_j,X_i](x) \right), 
\end{equation} 
for $x\in\Sigma_c$. Taking into account~\eqref{eq:xjxi} we achieve
\begin{equation}
 \sum_{i=1}^{n-r} X_j[\lambda_i] (x) X_i (x) =0 \mtext{for} 1\leq j\leq n+r,
\mtext{and}x\in\Sigma_c.
\end{equation} 
This implies that  $X_j[\lambda_i] (x) =\rmd \lambda_i(x)\cdot X_j(x)=0$ for
$1\leq j\leq n+r$ and $1\leq i\leq n-r$. In other words,  functions $\lambda_1,
\ldots, \lambda_{n-r}$ are constant on  $\Sigma_c$.
Summarising, if $ A \in \mathfrak{g}$,  then on $\Sigma_c$ we have
\begin{equation}
 v_A = \sum_{i=1}^{n-r}\lambda_i X_i, \mtext{where} \lambda_1, \ldots,
\lambda_{n-r}\in\C. 
\end{equation} 
Let us take another element $B \in \mathfrak{g}$. On $\Sigma_c$ we can write 
\begin{equation}
 v_B = \sum_{i=1}^{n-r}\gamma_i X_i, \mtext{where} \gamma_1, \ldots, 
\gamma_{n-r}\in\C. 
\end{equation}
Let us calculate the Lie bracket of  $v_A$ and $v_B$   on $\Sigma_c$. We have
\begin{equation*}
 [v_A,v_B]=[ \sum_{i=1}^{n-r}\lambda_i X_i , \sum_{k=1}^{n-r}\gamma_k X_k  ] =
\sum_{i,k=1}^{n-r}\lambda_i \gamma_k [ X_i,X_k] =0,
\end{equation*}
because $X_1,\ldots, X_{n-r}$ are commuting vector fields.  Thus, by
Proposition~\ref{prop:com}, 
$A$ and $B$ commute. In this way we showed that $\mathfrak{g}$ is Abelian. 
\end{proof}

Now we proceed to the proof of Theorem~\ref{thm:my}.  Assume that the system is
integrable in the non-commutative sense in an open connected neighbourhood $U$
of $\Gamma$ with the holomorphic independent first integrals $F_1, \ldots, F_k$,
and let $2r = \rank [a_{ij}(y)]$ where $y=F(\varphi(t))$. 
Condition~\eqref{eq:dims} implies that $ r = k -n$. 

 If $F$ is a holomorphic first integral of the considered system, then the
variational equations~\eqref{eq:svar} have a polynomial first integral $f$. It
is the first non-vanishing term in the Taylor expansion of $F$ around $\Gamma$. 
More precisely, for $p\in\Gamma$ first integral $f$ is a polynomial function in
$T_pM$ which depends holomorphically on $p$. Thus, the first integrals
$F_1,\ldots, F_k$  give rise  to first integrals $f_1, \ldots, f_k$ of the
variational equations. By the Ziglin Lemma, see section~1.3 in
\cite{Churchill:96::b} or Lemma III.1.7 in \cite{Audin:01::c}, we can assume
that $f_1, \ldots, f_k$ are functionally independent. 
Moreover,  let us fix  $p\in\Gamma$ and let $V= T_pM$, $\Omega=\omega(p)$.
Equations~\eqref{eq:FiFj} imply that on $(V,\Omega)$ we have
\begin{equation*}
 \{f_i,f_j\}(x)=\alpha_{ij}(f_1(x),\ldots,f_k(x) )\mtext{for}1\leq i,j\leq k,
\mtext{and} x\in V,
\end{equation*}
for certain polynomial functions $\alpha_{ij}$ in $k$ variables. To see this, it
is enough to expand  both sides of equations~\eqref{eq:FiFj} around point $p$
and compare the lowest order terms. Note also that
$\rank [\alpha_{ij}]\leq 2r$. 

From \cite{Morales:01::b1,Morales:99::c} we know that the differential Galois group $\mathscr{G}$ of the variational
equations~\eqref{eq:svar} is a subgroup of
$\mathrm{Sp}(2n,\C)$ and that $f_1,\ldots,f_k$ are invariants of this group, see also 
Lemmas~III.2.3 and III.1.13 in \cite{Audin:01::c}. 

Applying the Lie-Cartan theorem to polynomial functions $f_1, \ldots, f_k$ we
obtain new independent holomorphic functions $h_1, \ldots, h_k$ that are also invariants of the differential Galois group of the
variational equations~\eqref{eq:svar}. Thus, among them 
there is at least $n-r$ functions, let us say $h_1, \ldots, h_{n-r}$, which commute with
all $h_i$ for $1\leq i\leq k$.

Thus, the Lie algebra $\mathfrak{g}$ of $\mathscr{G}$ preserves all $h_1,
\ldots, h_k$. Hence, by Lemma~\ref{lem:cle}, $\mathfrak{g}$ is Abelian and this
means that the identity component $\mathscr{G}^\circ$ of  $\mathscr{G}$ is
Abelian. This finishes our proof of Theorem~\ref{thm:my}. 

\begin{remark}
 Let us note that using Ziglin lemma from the holomorphic first integrals $h_1,\ldots,h_k$ it is possible to construct $k$ algebraically independent polynomial invariants of the Galois group  and this fact can be important in applications. Furthermore, from this it is possible to give an alternative proof of Lemma~ \ref{lem:cle} in the lines of the Morales-Ramis original papers (using Proposition~8 of \cite{Morales:01::b1} or Proposition~3.4 of \cite{Morales:99::c}) by assuming that the first integrals are polynomials (instead of holomorphic) and algebraically independent (instead of functionally independent, although in this case both concepts are equivalent \cite{Churchill:96::b}).
\end{remark}

\section*{Acknowledgements}
We would like to thank professor Jean-Marie Strelcyn who pointed our attention to important work \cite{MR2120177}. Thanks are also due to Juan~J. Morales~Ruiz for his useful comments.

For the second author this research has been partially supported by  the
European Community project GIFT  (NEST-Adventure Project no. 5006) and by Projet
de l'Agence National de la Recherche  "Int\'egrabilit\'e r\'eelle et complexe en
m\'ecanique
hamiltonienne" N$^\circ$~JC05$_-$41465.


 \newcommand{\noopsort}[1]{}\def\cprime{$'$}

\end{document}